\documentclass[12pt]{iopart}

\begin{document}
\title{Symmetries and Algebraic Methods in Neutrino Physics}
\author{ A.~B. Balantekin}
\address{Department of Physics, University of Wisconsin - Madison, Wisconsin 53706 USA}
\ead{baha@physics.wisc.edu}

\begin{abstract}
Symmetry properties associated with neutrino propagation with or without a background of other particles, including neutrinos, is reviewed. 
The utility of symmetries is illustrated with examples chosen from the see-saw mechanism and both matter-enhanced and collective neutrino oscillations. The role of symmetries in neutrino astrophysics is highlighted. 

\end{abstract}
%
\vspace{2pc}

\submitto{\JPG}

\maketitle

\vskip 1.3cm

\section{Introduction}
\label{Section: Introduction}

Neutrinos are possibly the most abundant particles in the Universe. The number of neutrinos leftover from the Big Bang is sufficient enough to close the Universe were they slightly more massive. They are produced during the stellar evolution and even more copiously following supernovae explosions. According to the Standard Model of particle physics their lifetimes are longer than the age of the Universe. Hence, once they are formed they hang on, forming a diffuse supernova neutrino background. Despite this abundance, after they were first proposed by Pauli,  it took a long time before they can be detected and many of their properties can be measured. Today, with the high-statistics reactor neutrino experiments, neutrino physics has become a precision science. As a result of solar, atmospheric, accelerator, reactor and geophysical experiments we know that there are three active flavors of neutrinos which are not massless. Furthermore weak interaction eigenstates are not the same as mass eigenstates, instead those two eigenstates are related by a unitary transformation. Experimenters were able to measure the differences between squares of those masses and the mixing angles, but not the CP-violating phases, in this unitary transformation. 

Symmetries and conservation laws play a key role in understanding many physical phenomena and neutrino physics is no exception. The purpose of this article is to briefly review symmetries especially of neutrino oscillations and neutrino transport. In the next section we show that 
a symmetry associated with the SO(5) algebra and its subalgebras connects Dirac and Majorana masses, see-saw mechanism, and 
Pauli-G\"ursey transformation. In section III, we present the algebraic structure of neutrino mixing with three active and an undetermined number of sterile neutrinos. Section IV covers symmetries of collective neutrino oscillations which take place when a large number of neutrinos are present as well as the role of these symmetries in approximate solutions such as the mean-field approaches. A more technical presentation of the symmetries of collective neutrino oscillations is placed in an Appendix. Section V describes why these symmetries impact phenomena in neutrino astrophysics.  Finally Section VI contains a brief conclusion. 

\section{Symmetries of neutrino mass}

\subsection{Dirac versus Majorana neutrinos}

The field associated with a free Dirac neutrino is a spinor with four independent components. Hence a Dirac neutrino, which carries a conserved lepton number $L=+1$ is distinct from a Dirac antineutrino, which carries a lepton number $L=-1$. In contrast, the field associated with a free Majorana neutrino is a spinor is equal to its charge conjugate, hence carries no conserved lepton number. Whether neutrinos are Dirac or Majorana particles is not yet established. This is because most of the neutrinos which are experimentally accessible are 
ultrarelativistic  and the weak interactions are left-handed, resulting in helicity being an excellent substitute for the lepton number. For a recent review of the neutrino properties the reader is referred to Ref. \cite{Balantekin:2018azf}. 

The Lagrangian for the Dirac mass term is
\begin{equation}
{\cal L} = - m_D \overline{\psi} \psi = - m_D \left( \overline{\psi_R} \psi_L + {\rm h.c.} \right).  
\end{equation}
For charged leptons the Dirac mass term is the only possibility to describe the interaction between left- and right-handed spinors. There is yet another possibility for a mass term. The charge conjugate of a, for example, left-handed spinor $\psi_L$ is right-handed. Hence a term like 
$\overline{\psi_L} \psi^c_L$ would also be a mass term. Such a term was introduced by E. Majorana who found a real representation of Dirac matrices \cite{Majorana:1937vz}. However a Majorana mass term violates charge conservation, hence it is only possible for neutral fermions, such as neutrinos. The Lagrangian for the Majorana mass term composed of left-handed spinors is 
\begin{equation}
{\cal L} = - \frac{m_L}{2} \left( \overline{\psi_L} \psi_L^c + {\rm h.c.} \right), 
\end{equation}
where the factor $1/2$ is introduced to avoid double counting. 

It is instructive to write the static Dirac Hamiltonian 
\begin{equation}
H_D = \int d^3x \overline{\psi} ( -i \vec{\gamma} \cdot \nabla + m_D ) \psi 
\end{equation}
in terms of the massless fields 
\begin{equation}
\hspace*{-2.5cm}
\psi_L ( \mathbf{x}, t) = \int d{\cal P} \left[ a(\mathbf{p} , h=-1) u(\mathbf{p} , h=-1) e^{-ip.x} + b^{\dagger} (\mathbf{p} , h=+1) v( \mathbf{p} , h=+1) e^{ip.x} \right] ,
\end{equation}
where $d{\cal P} = d^3\mathbf{p} / 2E (2\pi)^3$ is the Lorentz invariant measure, $u$ and $v$ are the Dirac spinors in the helicity basis, $a$ and $b$ are the associated creation-annihilation operators, 
and
\begin{equation}
\hspace*{-2.5cm}
\psi_R ( \mathbf{x}, t) = \int d{\cal P} \left[ a(\mathbf{p} , h=+1) u(\mathbf{p} , h=+1) e^{-ip.x} + b^{\dagger} (\mathbf{p} , h=-1) v( \mathbf{p} , h=-1) e^{ip.x} \right] .
\end{equation}
One gets
\begin{eqnarray}
H_D &=& \sum_h \int d{\cal P} \>\> \left\{ \mathbf{p} \left[  a^{\dagger} (\mathbf{p} , h) a(\mathbf{p} , h)  - b (-\mathbf{p} , h) 
b^{\dagger} (- \mathbf{p} , h) \right]  \right. \nonumber \\
&+&  m_D \left. \left[ a^{\dagger} (\mathbf{p} , h) b^{\dagger} (- \mathbf{p} , h) +
b( - \mathbf{p} , h)a (\mathbf{p} , h) \right]  \right\}   .
\label{HD}
\end{eqnarray} 
Defining
\begin{equation}
\label{Q+}
Q_+(\mathbf{p},h) =  a^{\dagger} (\mathbf{p},h) b^{\dagger} (-\mathbf{p},h) = \left( Q_- (\mathbf{p},h)\right)^{\dagger} ,
\end{equation}
and 
\begin{equation}
\label{Q0}
Q_0 (\mathbf{p}, h) = \frac{1}{2}  \left[ a^{\dagger} (\mathbf{p},h) a (\mathbf{p},h) - b(-\mathbf{p},h) b^{\dagger} (-\mathbf{p},h) \right]
\end{equation}
the Dirac Hamiltonian of Eq. (\ref{HD}) can be written as 
\begin{equation}
\label{HD2}
H_D = \sum_h \int d{\cal P}  \left\{ 2 \mathbf{p} \> Q_0(\mathbf{p},h)  +m_D \left[ Q_+ (\mathbf{p},h) + Q_- (\mathbf{p},h) \right] \right\} .
\end{equation}
The Hamiltonian in Eq. (\ref{HD2}) is mathematically analogous to the linearized BCS Hamiltonian of superconductivity except that the 
operators $Q_+, Q_-$ and $Q_0$ do not satisfy the SU(2) commutation relations. However their integrals 
\begin{equation}
{\cal Q}_{+,-,0} = \sum_h \int d{\cal P} \>  Q_{+,-,0} (\mathbf{p},h) 
\end{equation}
satisfy the commutation relations of the SU(2) algebra:
\begin{equation}
\label{Qalgeb}
[{\cal Q}_+, {\cal Q}_- ] = 2 {\cal Q}_0 , \>\> [{\cal Q}_0, {\cal Q}_+] = {\cal Q}_+ , \>\> [{\cal Q}_0, {\cal Q}_-] = - {\cal Q}_- . 
\end{equation}
Hence the mass term in the field theory can be considered as resulting from a pairing interaction between left- and right-handed components of a fundamental spinor field.  In fact the mathematical analogy between particle mass and the pairing gap in the BCS theory of superconductivity was already remarked by Nambu and Jona-Lasinio in their seminal paper \cite{Nambu:1961tp} . 
(For a historical account see Ref. \cite{Nambu:2011zz}). 

The Hamiltonian in Eq. (\ref{HD}) can be diagonalized by a Bogoliubov transformation
\begin{eqnarray} 
A (\mathbf{p},h) &=& \cos \alpha \> a(\mathbf{p},h) - \sin \alpha \> b^{\dagger} (- \mathbf{p},h) , \\
B^{\dagger} (- \mathbf{p},h) &=& \sin \alpha \> a(\mathbf{p},h) + \cos \alpha \> b^{\dagger} (- \mathbf{p},h) 
\end{eqnarray}
where
\begin{equation}
\cos 2 \alpha = \frac{\mathbf{p}}{E}, \>\>\> \sin 2 \alpha = \frac{m_D}{E}
\end{equation}
with 
\begin{equation}
E = \sqrt{\mathbf{p}^2 + m_D^2}. 
\end{equation}
This result illustrates the fact that as the neutrino energy increases the left-handed component of the neutrinos increases as $\mathbf{p}/E$ whereas the right-handed component decreases as $m/E$. (Even though the calculation above is carried out for Dirac neutrinos, one gets the same conclusion for Majorana neutrinos). Similarly as the antineutrino energy increases the right-handed component of the antineutrinos increases as $\mathbf{p}/E$ whereas the left-handed component decreases as $m/E$.

\subsection{Algebraic structure of the see-saw mechanism}

Why the neutrino masses are much smaller than the masses of charged leptons is an unsolved puzzle. 
In the simplest version of the Standard Model there are no right-handed neutrino fields, hence neutrinos are taken to be massless. However, introducing a right-handed neutrino field, scalar under the the electroweak symmetry group $SU(2)_W \times U(1)$, it is possible to write down a Dirac mass term for neutrinos. In effective field theories at lower energies beyond the Standard Model physics is described by local operators, successively increasing in mass dimension, but suppressed by the powers of the energy scale of new physics:
\begin{equation}
\label{eff}
{\mathcal L} = {\mathcal L}_{\rm SM} + \frac{C^{(5)}}{\Lambda} {\mathcal O}^{(5)} +\sum_i 
\frac{C^{(6)}_i}{\Lambda^2} {\mathcal O}^{(6)}
+\sum_i \frac{C^{(7)}_i}{\Lambda^3} {\mathcal O}^{(7)} + \cdots ,
\end{equation}
where $\Lambda$ is the scale of new physics, ${\mathcal O}^{(n)}_i$ are the possible operators of dimension $n$, and $C^{(n)}$ are strengths of their contribution to the Lagrangian. 
There is only a single mass-dimension five correction to the Standard Model Lagrangian  \cite{Weinberg:1979sa} and it provides a Majorana mass for neutrinos. 

Combining Dirac and Majorana masses, the see-saw mechanism provides one possible explanation for the smallness of the neutrino masses \cite{Yanagida:1980xy}. One starts with the mass term of the Lagrangian containing both Dirac and Majorana Masses: 
\begin{eqnarray}
{\cal L}_m &=& - \frac{1}{2} \left[ m_L \overline{\psi^c_L} \psi_L + m_R \overline{\psi_R} \psi^c_R + 2 m_D \overline{\psi_R} \psi_L + {\rm h.c.}\right] \nonumber \\ &=&
 - \frac{1}{2} 
\left( \overline{\psi^c_L} ,  \overline{\psi_R} \right) \left(
\begin{array}{cc}
 m_L & m_D   \\
m_D  &  m_R   
\end{array}
\right)
\left(
\begin{array}{c}
  \psi_L   \\
\psi^c_R
\end{array}
\right) + {\rm h.c.} .
\label{allmasses}
\end{eqnarray}
Diagonalizing the mass matrix this can be written as 
\begin{eqnarray}
\label{lightheavy}
{\cal L}_m &=& - \frac{1}{2} \left[ m_+ + \sqrt{ m_-^2 + m_D^2} \right] \left( \cos  \varphi \> \psi_L 
+ \sin \varphi \> \psi^c_R \right) \nonumber \\
&-& \frac{1}{2} \left[ m_+ - \sqrt{ m_-^2 + m_D^2} \right] \left( - \sin  \varphi \> \psi_L 
+ \cos \varphi \> \psi^c_R \right) + {\rm h.c.}
\end{eqnarray}
where
\begin{equation}
\label{varphi}
\hspace*{-0.7cm}
\cos \varphi = \frac{1}{\sqrt{2}} \left( 1 + \frac{m_-}{\sqrt{m_-^2 +  m_D^2}} \right)^{1/2}, \>\> \>\>
\sin \varphi = \frac{1}{\sqrt{2}} \left( 1 - \frac{m_-}{\sqrt{m_-^2 +  m_D^2}} \right)^{1/2}
\end{equation}
with 
\begin{equation}
m_{\pm} = \frac{m_L \pm m_R}{2} .
\end{equation}
The basic idea of the see-saw mechanism is very simple. In a scenario where $m_L=0$ and $m_R \gg m_D$, Taylor expanding the quantities in Eq. (\ref{lightheavy}) in powers of $m_D/m_R$ and keeping the lowest order terms one obtains one light neutrino with mass $\sim m_D^2/m_R$ associated with the state $\sim \psi_L - (m_D/m_R) \psi^c_R$ and one heavy neutrino with mass $\sim m_R$, associated with the state 
$\sim (m_D/m_R) \psi_L + \psi^c_R$. There are many variants of the see-saw models in the literature. 

We next explore, following Ref.  \cite{Balantekin:2000qt},  the underlying algebraic basis of the see-saw mechanism. First one can show that the three sets of operators, 
\begin{equation}
\label{dirac}
D_- = \int d^3{x}(\bar{\psi}_R\psi_L) = D_+^\dagger, \> 
D_0=\frac{1}{2}\int d^3{x}(\psi_L^{\dag}\psi_L-\psi_R^{\dag}\psi_R) ,
\end{equation}
\begin{equation}
\label{leftmaj}
L_+ = \frac{1}{2}\int d^3{x}(\bar{\psi}_L\psi_L^c) = L_-^\dagger, 
L_0=\frac{1}{4}\int d^3{x}(\psi_L^{\dag}\psi_L-\psi_L\psi_L^{\dag}),
\end{equation}
and 
\begin{equation}
\label{rightmaj}
R_+=\frac{1}{2}\int d^3{x} (\bar{\psi_R^c}\psi_R) = R_-^\dagger,  
R_0=\frac{1}{4}\int d^3{x}(\psi_R \psi_R^{\dag}-\psi_R^{\dag}\psi_R) 
\end{equation}
generate three distinct  SU(2) algebras. We designate them as SU(2)$_D$, SU(2)$_L$ and SU(2)$_R$, respectively. In Eqs. (\ref{dirac}), ({\ref{leftmaj}), and (\ref{rightmaj}), $\psi_L$ and $\psi_R$ are the left- and right-handed components of the same spinor field and the superscript $c$ indicates their charge conjugates as before. The SU(2)$_D$ algebra is the same SU(2) algebra given in Eq. (\ref{Qalgeb}), but in rotated basis, i.e. the generators $D_{\pm,0}$ are linear combinations of the generators ${\cal Q}_{\pm,0}$. 

Note that the algebras SU(2)$_L$ and SU(2)$_R$ are mutually commuting. However  SU(2)$_D$ does not commute with them, in particular $D_0$ is not an independent operator, but the sum of $L_0$ and $ R_0$. The most general neutrino mass Hamiltonian (\emph{not}  the Hamiltonian density)\footnote{This is the Hamiltonian associated with the Lagrangian density in Eq. (\ref{allmasses}) up to an overall factor.} can be written in terms of the operators listed above: 
\begin{equation}
\label{massham}
H_{m}=m_D(D_++D_-)+m_L(L_++L_-)+m_R(R_++R_-), 
\end{equation}
where $m_D$ is the Dirac mass with $m_L$ and $m_R$ being Majorana masses associated with left- and right-handed spinors, respectively. Introducing two more operators, 
\begin{equation}
\label{As}
A_+ = \int d^3{x} \left[ -\psi_L^T C\gamma_0\psi_R \right], \> 
 A_-= \int
d^3{x} \left[ \psi_R^{\dag} \gamma_0 C (\psi_L^{\dag})^T\right], 
\end{equation}
one can show that the operators $D_{\pm},L_{\pm},R_{\pm}, L_0, R_0$ and $A_{\pm}$ are the ten generators of the SO(5) algebra. 

To establish the connection to the see-saw mechanism we need to identify yet another $SU(2)$ subalgebra of SO(5). The operators $A_{\pm}$ along with $A_0 \equiv R_0 - L_0$ also generate an SU(2) algebra. As we show below this fourth SU(2) algebra is related to the 
Pauli-G\"{u}rsey transformation \cite{PG},  
\begin{equation}
\psi\rightarrow\psi^{\prime}=a\psi+b\gamma_5\psi^c,  \>\> |a|^2+|b|^2=1 .
\end{equation} 
Under the SU(2) group associated with this algebra,
\begin{equation}
\label{pgsu2}
\hat{U}=e^{-\tau^* A_-}e^{-\log(1+|\tau|^2)A_0}e^{\tau A_+}e^{i\varphi
A_0}, 
\end{equation} 
the field $\psi$ transforms as 
\begin{equation}
\psi \rightarrow \psi' =
\hat{U}\psi\hat{U}^{\dag}=\frac{e^{i\varphi/2}}{\sqrt{1+|\tau|^2}}
[\psi-\tau^* \gamma_5 \psi^c] ,
\end{equation}
which is a Pauli-G\"{u}rsey transformation with
\begin{equation}
a=\frac{e^{i\varphi/2}}{\sqrt{1+|\tau|^2}} \;\; ,\;\; b=\frac{-\tau^*
e^{i\varphi/2}}{\sqrt{1+|\tau|^2}} .
\end{equation} 
It can easily be shown that under the transformation of Eq. (\ref{pgsu2}) with $\tau = \tan \varphi$ where $\varphi$ is given by the Eq. (\ref{varphi}), 
the Hamiltonian of Eq. (\ref{massham}) transforms as 
\begin{eqnarray}
H_m \rightarrow H_m' = \hat{U} H_m \hat{U}^{\dagger} &=& m_+ \left[ \left( L_+ + L_- \right) + \left( R_+ + R_- \right) \right] \nonumber \\
&-& \left[ m_-^2 + m_D^2 \right]^{1/2} \left[ \left( L_+ + L_- \right) + \left( R_+ + R_- \right) \right] .
\end{eqnarray}
Making the choice  $m_L=0$ and $m_R \gg m_D$ as before, one obtains the result 
\begin{equation}
H_m' \sim m(R) \left( R_+ + R_- \right) - \frac{m_D^2}{m_R} \left( L_+ + L_- \right) .
\end{equation}
Hence the Pauli-G\"ursey SU(2)  rotation generated by the operators $A_{\pm}, A_0$ produces a see-saw type transformation.

\section{Neutrino mixing and oscillations}

Experimental efforts during the last several decades firmly established that there are three active neutrino flavors which participate in the electroweak interactions, that at least two of those flavors are massive, and that the mass eigenstates and electroweak eigenstates do not coincide, but are related by a linear transformation:
\begin{equation}
| \nu_f \rangle = \sum_i U_{fi} | \nu_i \rangle ,
\end{equation}
where $f$  and $i$ are the flavor and mass basis indices, respectively. 
It is not yet clear if this transformation $U$ is unitary with only the three active neutrinos, or if there are additional mass eigenstates which do not directly take part in the weak interactions, but nevertheless mix with the active flavors which do. 
If there are no such sterile flavors the $3 \times 3$ mixing matrix for the active neutrinos would be a unitary matrix. At the moment there are hints, but no direct experimental evidence for such sterile states. 

If there are $N$ flavors (active plus sterile), the mixing matrix will be an $N \times N$ unitary matrix with determinant one, i.e. the fundamental representation of the group SU(N).  Such matrices are parameterized by $N^2 -1$ independent real parameters. In general a given element of the U(N) group can be written as a product of $N(N-1)/2$ distinct and non-commuting SU(2) rotations and a diagonal matrix of pure phases. Each of those SU(2) rotations are parameterized by one Euler angle and one phase. Hence an $N\times N$ mixing matrix can be written using 
$N(N-1)/2$ Euler angles and 
\begin{equation}
\left[\frac{1}{2} N(N-1) \right] + N -1
\end{equation}
phases\footnote{Taking all but one of the entries in two of the rows of the mixing matrix to be proportional results in the decoupling of one of the mass eigenstates from the mixing matrix \cite{Balantekin:1999dx}. Such restrictions may help model building \cite{Barger:1998ta,Harrison:2002et}.}.  The $-1$ factor above accounts for the overall phase set to zero to impose unitarity; this is required to ensure that physics does not depend on the choice of basis).
Some of these phases can be absorbed into the definition of neutrino states. The number of remaining phases depend on whether neutrinos are Dirac or Majorana particles since Majorana fermions need to be self charge-conjugate.  For three active and no sterile flavors a commonly used parameterization is 
\begin{eqnarray}
\hspace*{-0.9cm}
U = \left(
\begin{array}{ccc}
 1 & 0  & 0  \\
  0 & C_{23}   & S_{23}  \\
 0 & -S_{23}  & C_{23}  
\end{array}
\right)
\left(
\begin{array}{ccc}
 C_{13} & 0  & S_{13} e^{-i\delta_{CP}}  \\
 0 & 1  & 0  \\
 - S_{13} e^{i \delta_{CP}} & 0  & C_{13}  
\end{array}
\right) 
\left(
\begin{array}{ccc}
 C_{12} & S_{12}  & 0  \\
 - S_{12} & C_{12}  & 0  \\
0  & 0  & 1  
\end{array}
\right) \nonumber \\
\hspace*{0.7cm}
\times 
\left(
\begin{array}{ccc}
 1 & 0  & 0  \\
 0 & e^{i\alpha_1/2}  & 0  \\
0  & 0  &   e^{i\alpha_2/2}
\end{array}
\right), 
\label{matrix}
\end{eqnarray}
where $C_{ij} = \cos \theta_{ij}$, $S_{ij} = \sin \theta_{ij}$, $\delta_{CP}$ is the CP-violating phase and $\alpha_{1,2}$ are the Majorana phases. 

Since different mass eigenstates propagate  with different velocities, the detection amplitude of any flavor will oscillate as neutrinos travel. The 
propagation is governed by the equation 
\begin{equation}
\label{oscillation}
i \frac{\partial}{\partial t} | \nu_f \rangle = \left[ U_{ f i} \left( p + \frac{m_i^2}{2p} \right) \delta_{ij} U^{\dagger}_{j f'} + v_f \delta_{f f'} \right] | \nu_{f'} \rangle 
\end{equation}
where $m_i$ is the mass associated with the i{\it th}  mass eigenstate, p is the common momenta, $v_e = V_C + V_N$, $v_{\mu} = 
v_{\tau} = V_N$ with 
\begin{equation}
\label{charged}
V_C = \pm \sqrt{2} G_F \left[ N_{e^-} (x) - N_{e^+} (x) \right]   
\end{equation}
where the plus sign refers to electron neutrinos, the minus sign refers to electron antineutrinos 
and 
\begin{equation}
\label{neutral} 
V_N = - \frac{1}{\sqrt{2}} G_F N_n (x) 
\end{equation} 
where $N_e$ and $N_n$ are the electron and neutron densities, respectively, of the background in which neutrinos travel through. $V_C$ and $V_N$ result from coherent forward scattering of neutrinos from the background particles, calculated using tree-level diagrams. In writing Eqs. (\ref{charged}) and (\ref{neutral}) the background is assumed to be static and locally charge-neutral: the proton density is taken to be the same as $N_e$. A term proportional to the identity contributes the same overall phase to all the flavors and do not effect the oscillations. Consequently if there are no sterile flavors, 
$N_n$ drops out of the oscillation amplitudes, but it contributes if there are one or more sterile flavors since $v_f = 0$ for those additional flavors. 

Oscillation experiments determine either the survival probability of the original flavor (``disappearance" experiments) or the appearance of a new flavor (``appearance" experiments). The Majorana phases do not contribute to the either probability. CP-violating phases drop out from some, but not all, of the oscillation probabilities. If they are set to zero the neutrino mixing matrix becomes an element of the SO(N) algebra for N flavors\footnote{See the discussion in the Appendix of Ref. \cite{Balantekin:2013sda}.}. But the diagonal term between the two mixing matrices as well as the term $v_f \delta_{f f'}$ in Eq. (\ref{oscillation}) belong to the coset SU(N)/SO(N) keeping neutrino propagation still an SU(N) evolution problem even when CP-violating phases vanish. 

It is convenient to write Eq. (\ref{oscillation}) in the form 
\begin{equation}
i \frac{\partial}{\partial t} | \nu_f \rangle =H_{f f'} | \nu_{f'} \rangle 
\end{equation}
where 
\begin{equation}
H_{f f'} = \left[ U_{ f i} \left( p + \frac{m_i^2}{2p} \right) \delta_{ij} U^{\dagger}_{j f'} + v_f \delta_{f f'} \right] .
\end{equation}
Introducing the combinations 
\begin{equation}
\label{tildemu}
\tilde{\nu}_{\mu} = \cos \theta_{23}\>  \nu_{\mu} - \sin \theta_{23} \> \nu_{\tau}, 
\end{equation}
and 
\begin{equation}
\tilde{\nu}_{\tau} = \sin \theta_{23}\>  \nu_{\mu} + \cos \theta_{23} \> \nu_{\tau} 
\end{equation}
for three active flavors Eq. (\ref{oscillation}) takes the form 
\begin{equation}
i \frac{\partial}{\partial t}
\left(
\begin{array}{c}
  | \nu_e \rangle  \\
 | \tilde{\nu}_{\mu} \rangle  \\
  | \tilde{\nu}_{\tau} \rangle
\end{array}
\right) = \tilde{H} 
\left(
\begin{array}{c}
  | \nu_e \rangle  \\
 | \tilde{\nu}_{\mu} \rangle  \\
  | \tilde{\nu}_{\tau} \rangle
\end{array}
\right) 
\end{equation}
where
\begin{equation}
\tilde{H} = \left[ \tilde{T} 
\left(
\begin{array}{ccc}
m_1^2/2p  &  0 & 0  \\
 0 & m_2^2/2p   & 0  \\
0  & 0  &  m_3^2/2p 
\end{array}
\right) \tilde{T}^{\dagger} + \left(
\begin{array}{ccc}
V_c  &  0 & 0  \\
 0 & 0   & 0  \\
0  & 0  &  0 
\end{array}
\right) 
\right]
\end{equation}
with
\begin{equation}
\tilde{T} = \left(
\begin{array}{ccc}
 C_{13} & 0  & S_{13} e^{-i\delta}  \\
 0 & 1  & 0  \\
 - S_{13} e^{i \delta} & 0  & C_{13}  
\end{array}
\right) 
\left(
\begin{array}{ccc}
 C_{12} & S_{12}  & 0  \\
 - S_{12} & C_{12}  & 0  \\
0  & 0  & 1  
\end{array}
\right) .
\end{equation}
In writing the equations above an overall phase is ignored. One can show that the dependence of the Hamiltonian $\tilde{H} (\delta)$ on the 
CP-violating phase can be factored out \cite{Balantekin:2007es}:
\begin{equation}
\label{factorout} 
\tilde{H} (\delta) = S^{\dagger}(\delta) \tilde{H} (\delta = 0) S (\delta)
\end{equation}
where
\begin{equation}
S(\delta) = 
\left(
\begin{array}{ccc}
 1 & 0  &   0\\
0  & 1  & 0  \\
0  &0   & e^{i\delta}  
\end{array}
\right) .
\end{equation}
Using this factorization one can relate survival or appearance probabilities for the two cases with $\delta =0$ and $\delta \neq 0$. It should be emphasized that this factorization is valid only for three active flavors and it no longer holds if there are one or more sterile neutrinos which mix with the active ones. 

Since an overall phase does not impact oscillations, only the two differences of the neutrino masses, $\delta m_{21}^2 = m_2^2 - m_1^2$ and 
 $\delta m_{32}^2 = m_3^2 - m_2^2$ contribute to the oscillation amplitudes. Similarly if one ignores the loop corrections \cite{Botella:1986wy} 
 to the Eqs. (\ref{charged}) and (\ref{neutral}) and excludes sterile neutrinos, only electron density contributes to these amplitudes. Experimentally these two mass differences were found to differ by two orders of magnitude. This separation of scales 
motivates a two-flavor description of neutrino oscillations in matter. A typically encountered form in the literature is 
\begin{equation}
\label{deviation}
i \frac{\partial}{\partial t} 
\left(
\begin{array}{c}
 | \nu_e \rangle  \\
 | \nu_x \rangle
\end{array}
\right)
=    \left(
\begin{array}{cc}
 \frac{G_F N_e}{\sqrt{2}} -  \frac{\delta m^2}{4E} \cos 2 \theta  &   \frac{\delta m^2}{4E} \sin 2 \theta   \\
 \frac{\delta m^2}{4E} \sin 2 \theta  &   -  \frac{G_F N_e}{\sqrt{2}} + \frac{\delta m^2}{4E}  \cos 2 \theta
\end{array}
\right)
\left(
\begin{array}{c}
  | \nu_e \rangle  \\
  | \nu_x \rangle
\end{array}
\right) ,
\end{equation}
where $\nu_x$ can be taken as a linear combination of $\nu_{\mu}$ and $\nu_{\tau})$. 
If $\theta_{13}$ were exactly zero, then the above equation would be exact with $\nu_x$ given by Eq. (\ref{tildemu}) with $\theta = \theta_{12}$ and $\delta m^2 = \delta m_{21}^2$ \cite{Balantekin:2011ta}. 

\section{Collective neutrino oscillations}

In certain astrophysical environments such as core-collapse supernovae and merging of binary neutron stars a very large number of neutrinos 
are present. For example in a core-collapse supernova almost all the gravitational binding energy of the pre-supernova star is deposited in the 
proto-neutron star, which cools by emitting neutrino-antineutrino pairs. In such situations even the average energies of different flavors are  different since only electron neutrinos participate in charged-current weak interactions. In these environments it is no longer possible to ignore the contribution of the neutrino-neutrino scattering to neutrino propagation. Neutrino transport then becomes a many-body problem, hence it is more transparent to use the many-body language. To this end we introduce creation and annihilation operators for neutrinos and antineutrinos.  
It is more instructive to first consider a many-neutrino system containing only two flavors of neutrinos, which we take to be the electron neutrino, $\nu_e$, and an unspecified flavor, $\nu_x$. Later we will add the third flavor and antineutrinos. 

Introducing the creation and annihilation operators for one neutrino 
with three momentum ${\bf p}$, we can write down the generators of an SU(2) algebra 
\cite{Balantekin:2006tg}: 
\begin{eqnarray}
J_+({\bf p}) &=& a_x^\dagger({\bf p}) a_e({\bf p}), \> \> \>
J_-({\bf p})=a_e^\dagger({\bf p}) a_x({\bf p}), \nonumber \\
J_0({\bf p}) &=& \frac{1}{2}\left(a_x^\dagger({\bf p})a_x({\bf p})-a_e^\dagger({\bf p})a_e({\bf p})
\right). \label{su2}
\end{eqnarray}
In fact there are as many SU(2) algebras as the number of different momenta, which commute with each other\footnote{To be mathematically rigorous one should specify the values of the momenta to be discrete following a box quantization so that one gets an SU(2) algebra instead of a current algebra.}.  
The sum of these operators over all possible values of momenta also 
generate a global SU(2) algebra. 
Using the operators in Eq. (\ref{su2}) 
the Hamiltonian for a neutrino propagating through matter takes the form  
\begin{equation}
\label{msw}
\hspace*{-2.5cm}
 H_{\nu} = \int d^3{\bf p} \frac{\delta m^2}{2p} \left[
\cos{2\theta} J_0({\bf p}) + \frac{1}{2} \sin{2\theta}
\left(J_+({\bf p})+J_-({\bf p})\right) \right] -  \sqrt{2} G_F \int d^3{\bf p} 
\> N_e \>  J_0({\bf p}).  
\end{equation}
In Eq. (\ref{msw}), the first integral represents the neutrino mixing and the second integral 
represents the neutrino forward scattering off the background matter.  Since different SU(2) algebras for different momenta commute with each other, the propagation of neutrinos with a given momentum is independent of the propagation of neutrinos carrying other momentum values. 
Note that $N_e$ in the second term of Eq. (\ref{msw}) is inside the momentum integral since electron densities encountered by neutrinos traveling in different directions can be different. If one substitutes the two-dimensional  Pauli matrices representation of SU(2) in Eq. (\ref{msw}) one obtains the Hamiltonian of Eq. (\ref{deviation}) for a given momentum. 
Neutrino-neutrino forward-scattering contributions are described by the Hamiltonian 
\begin{equation}
\label{nunu}
H_{\nu \nu} = \sqrt{2} \frac{G_F}{V} \int d^3{\bf p} \> d^3{\bf q} \>  (1-\cos\vartheta_{\bf pq}) \> {\bf
J}({\bf p}) \cdot {\bf J}({\bf q}) ,
\end{equation}
where $\vartheta_{\bf pq}$ is the angle between neutrino momenta {\bf p} and {\bf q} and V 
is the normalization volume.  The $(1-\cos\vartheta_{\bf pq}) $ term in the integral above 
ensures that neutrinos traveling in the same direction do not forward scatter off each other. Note that contributions from collisions are proportional to $G_F^2$ and can be ignored in a first approximation. 

The total Hamiltonian $H_{\nu} + H_{\nu \nu}$ describes collective neutrino oscillations. Note that the only term which contains the neutrino masses is the first term in Eq. (\ref{msw}). In particular in writing $H_{\nu \nu}$ neutrino masses are set to zero. 
Antineutrinos can be incorporated by introducing a second set of SU(2) algebras. 
There are excellent reviews of the astrophysics applications of the collective neutrino oscillations \cite{Duan:2009cd,Duan:2010bg}; those applications are beyond the scope of this article which focuses on symmetries and algebraic approaches. 

\subsection{Single-angle approximation and conserved quantities}

For pedagogical convenience in this subsection we take neutrino momenta to be discrete and replace the integrals with sums. 
When neutrino-neutrino interactions significantly contribute to neutrino propagation they are the dominant contribution to the Hamiltonian. Hence in most cases one can ignore the neutrino interactions with the background electrons.  In this approximation one can write the total 
Hamiltonian in a compact form as 
\begin{equation}
\label{multiangle}
H = \sum_p \omega_p \mathbf{B} \cdot \mathbf{J} (\mathbf{p}) + \sqrt{2} \frac{G_F}{V}  \sum_{p,q} (1-\cos\vartheta_{\bf pq}) \> {\bf
J}({\bf p}) \cdot {\bf J}({\bf q})  
\end{equation} 
where $\omega_p = \delta m^2/ 2p$ and the auxiliary vector $\mathbf{B}$ was introduced. The Hamiltonian in Eq. (\ref{multiangle}) 
can be written either in the mass or flavor basis; those two Hamiltonians are related by a unitary transformation. The second term is the same in both bases whereas the first term changes as
\begin{equation}
{\mathbf B}_{\rm mass} = (0,0, -1)
\end{equation}
or
\begin{equation}
{\mathbf B}_{\rm flavor} = (\sin 2 \theta,0, -\cos 2 \theta). 
\end{equation}
A further approximation is to replace the quantity $(1-\cos\vartheta_{\bf pq})$ by its average value over the ensemble; this is known as the single-angle approximation resulting in the Hamiltonian
\begin{equation}
\label{singleangle}
H = \sum_p \omega_p \mathbf{B} \cdot \mathbf{J} (\mathbf{p}) +  \mu \sum_{p,q,p \neq q}  {\bf
J}({\bf p}) \cdot {\bf J}({\bf q})  
\end{equation} 
where we defined $\mu = ( \sqrt{2} G_F/V)  \langle 1-\cos\vartheta_{\bf pq} \rangle$. The Hamiltonian in Eq. (\ref{singleangle}) is mathematically analogous to the BCS Hamiltonian and one can find its eigenvalues and eigenfunctions using a Bethe ansatz. The details of this approach is described in the Appendix.  One finds that the quantities \cite{Pehlivan:2011hp} 
\begin{equation}
h_p = {\mathbf B} \cdot {\mathbf J} +  \mu  \sum_{q,q\neq p} \frac{\mathbf{J} ({\mathbf p}) \cdot \mathbf{J} ({\mathbf q})}{\omega_p - \omega_q} 
\end{equation}
commute with one other:
\begin{equation}
[h_p, h_q] = 0. 
\end{equation}
It is  easy to show that the Hamiltonian of Eq. (\ref{singleangle}) can be written as 
\begin{equation}
H = \sum_p \omega_p h_p 
\end{equation}
demonstrating that $h_p$ are conserved quantities. One can form other combinations of $h_p$'s to construct different invariants. For example 
\begin{equation}
\sum_p h_p = {\mathbf B} \cdot {\mathbf J} 
\end{equation}
where ${\mathbf J} = \sum_p {\mathbf J}_p$. In addition to these invariants the overall SU(N) symmetry, where N is the number of neutrino flavors, shapes the neutrino energy spectra \cite{Duan:2008fd}.

It is possible to extend this analysis to three flavors of neutrinos and antineutrinos by introducing two sets of SU(3) algebras \cite{Pehlivan:2014zua} and construct the invariants. If one ignores the interaction between neutrino magnetic moments and external magnetic fields, one can also show that the CP-violating phase factors out of the evolution equations as it is demonstrated in Eq. (\ref{factorout}) for the case without neutrino-neutrino interactions \cite{Pehlivan:2014zua,Gava:2008rp}.  

\subsection{Mean Field Approximations}

Finding exact solutions to many-body Hamiltonians is exceedingly difficult, consequently approximations are introduced. In the mean-field approximation one approximates a two-body term written as a product of two operators $\hat{{\cal O}}_1$ and $\hat{{\cal O}}_2$ as 
\begin{equation}
\label{mpa}
\hat{{\cal O}_1} \hat{{\cal O}}_2 \sim \hat{{\cal O}_1} \langle \hat{{\cal O}}_2 \rangle + \langle \hat{{\cal O}_1} \rangle \hat{{\cal O}}_2 
- \langle \hat{{\cal O}_1} \rangle  \langle \hat{{\cal O}}_2 \rangle
\end{equation}
provided that the commutator $[\hat{{\cal O}_1} , \hat{{\cal O}}_2]$ is very small to begin with. Averages in Eq. (\ref{mpa}) are calculated with a suitably defined wavefunction. Clearly such a wavefunction should satisfy the condition $\langle \hat{{\cal O}_1} \hat{{\cal O}}_2  \rangle \sim \langle \hat{{\cal O}_1} \rangle \langle \hat{{\cal O}}_2 \rangle$. Product of a set of coherent states for the SU(2) algebras given in Eq. (\ref{su2}) and their antineutrino counterparts is commonly used as this wavefunction. In fact, one can rigorously show that the saddle-point approximation to the path integral describing evolution of the Hamiltonian in Eq. (\ref{multiangle}) provides a consistent mean-field if this path integral is written using the resolution of identity for these coherent states \cite{Balantekin:2006tg}. Another possibility to determine the mean field is to treat the problem using Bogoliubov, Born, Green, Kirkwood, and Yvon hierarchy method. In this method the exact density operator 
can be written in terms of a hierarchy of one-, two-, three-body,.. density operators where the mean field corresponds to the lowest order 
\cite{Volpe:2013uxl,Volpe:2015rla}. 

As mentioned before, in writing the Hamiltonian of Eq. (\ref{multiangle}) neutrino masses are ignored; all the neutrinos are taken to be left-handed and all the antineutrinos to be right-handed. The SU(2) algebras for neutrinos and antineutrinos commute and the product coherent state used to calculate the mean field gives a mean field which consists of either only neutrinos or only antineutrinos. It is possible to choose mean fields which contain both neutrinos and antineutrinos \cite{Serreau:2014cfa,Cirigliano:2014aoa}. Since such a mean field contains both left-handed and right-handed neutrino spinors, it has to be proportional to the neutrino mass. 

The mean-field Hamiltonian in the single-angle limit is 
\begin{equation}
H = \sum_p \omega_p \mathbf{B} \cdot \mathbf{J}_p + \mu {\mathbf P} \cdot {\mathbf J}
\end{equation}
where the mean field is given by ${\mathbf P} = \sum_p {\mathbf P}_p = \sum_p \langle {\mathbf J}_p \rangle$. Even though this Hamiltonian looks linear  in  the SU(2) generators one has to remember that ${\mathbf P}$ still contains information about these generators. To see that one can write the equation of motion
\begin{equation}
\frac{d}{dt} {\mathbf J}_p = ( \omega_p {\mathbf B} + \mu {\mathbf P}) \times {\mathbf J}_p .
\end{equation}
Clearly consistency of the formalism also requires the following equation to be satisfied:
\begin{equation}
\label{consistency}
\frac{d}{dt} {\mathbf P}_p = ( \omega_p {\mathbf B} + \mu {\mathbf P}) \times {\mathbf P}_p .
\end{equation}
Using Eq. (\ref{consistency}) one can show that the invariants in the mean field limit 
\begin{equation}
\langle h_p \rangle = {\mathbf B} \cdot {\mathbf P}_p + \mu \sum_{q, q \neq p} \frac{{\mathbf P}_p \cdot {\mathbf P}_q}{\omega_p - \omega_q}
\end{equation}
remains constant, i.e. $d\langle h_p \rangle /dt =0$. 

One interesting effect resulting from the collective neutrino oscillations is spectral swappings or splits, on the final neutrino energy spectra: at a particular energy these spectra are almost completely divided into parts of different flavors \cite{Raffelt:2007cb,Duan:2008za}. Spectral splits 
were originally observed in calculations using the mean-field approximation. However recently assuming the conditions are perfectly adiabatic so that the evolution of the eigenstates follow their variation with the interaction rate, it was shown in an exact calculation that an initial state which consists of electron neutrinos and antineutrinos of an orthogonal flavor develops a spectral split at exactly the same energy predicted by the mean field formulation \cite{Birol:2018qhx}.

\section{Connection to astrophysics}

Since neutrino interactions with ordinary matter are rather feeble, neutrinos can carry energy and entropy over astronomical distances without 
much impediment. Consequently they can have a very significant impact on astrophysical phenomena. In a main-sequence star neutrinos emitted from the nuclear reactions at the core provide an outward energy flux which counterbalances gravity. Once the star runs out of  fuel, during the resulting core-collapse capture of electrons on protons and nuclei produces a brief neutrino burst. The resulting proto-neutron star possesses almost all of the gravitational binding energy of the pre-supernova star: 10$^{53}$ ergs or 10$^{59}$ MeV. The quickest way 
to release this very large amount of energy is emitting it as neutrino-antineutrino pairs. As these neutrinos travel outward they impact many aspects of the core-collapse supernovae with their properties playing a salient role in controlling the dynamics. The average energy of these neutrinos is typically 10 MeV or so, hence altogether one has a total number of 10$^{57} \sim$ 10$^{58}$ neutrinos emitted, resulting in onset of collective neutrino oscillations. In addition, both the core-collapse supernovae and mergers of binary neutron stars are likely to be sites of various element formation scenarios. 

For many nucleosynthesis processes the yields are determined by the neutron-to-proton ratio in the relevant site. We next sketch how neutrinos control this ratio. We start with the expression for the 
mass fraction, $X_j$, of species of kind $j$: 
\begin{equation}
\label{nnn17}
X_j = \frac{N_j A_j}{ \sum_i  N_i A_i},
\end{equation}
where $N_j$ is the number of species of kind $j$ per unit volume, and $A_j$ is the atomic
weight of the $j$-th species. Then the number abundance of species $j$ relative to baryons, $Y_j$, is given by 
\begin{equation}
\label{nnn18}
Y_j = \frac{X_j}{A_j} = \frac{N_j}{\sum_i N_i A_i} .
\end{equation}
The electron fraction, $Y_e$, is the net number of electrons (number
of electrons minus the number of positrons) per baryon:
\begin{equation}
\label{nnn16}
Y_e = ( n_{e^-} - n_{e^+} ) / n_B ,
\end{equation}
where $n_{e^-}$, $n_{e^+}$, and $n_B$ are number densities of
electrons, positrons, and baryons, respectively. Using Eq. (\ref{nnn18}) it takes the form 
\begin{eqnarray}
\label{nnn19}
Y_e &=& \sum_j Z_j Y_j = \sum_i \left( \frac{Z_j}{A_j} \right) X_j
\nonumber \\
&=& X_p + \frac{1}{2} X_{\alpha} +  \sum_h \left( \frac{Z_h}{A_h}
\right) X_h ,
\end{eqnarray}
where $Z_j$ is the charge of the species of kind $j$, and  $X_p$, $X_{\alpha}$, and
 $X_h$, are the mass fractions of protons, alpha particles, and heavier nuclei (``metals'' in astronomy parlance), respectively. 
 
Primary reactions that control the neutron-to proton ratio is the capture
reactions on free nucleons
\begin{equation}\label{nnn11}
\nu_e + {\rm n}  \rightleftharpoons {\rm p}+ e^{-} ,
\end{equation}
and
\begin{equation}\label{nnn12}
\bar{\nu}_e + {\rm p} \rightleftharpoons {\rm n} + e^{+} .
\end{equation}
The rate of change of the number of protons is given by 
\begin{equation}
\label{nnn20}
\frac{dN_p}{dt} = - ( \lambda_{\bar{\nu}_e} + \lambda_{e^-} ) N_p
+  ( \lambda_{\nu_e} + \lambda_{e^+} ) N_n ,
\end{equation}
where $\lambda_{\nu_e}$ and $\lambda_{e^-}$ are the rates of the
forward and backward reactions in Eq. (\ref{nnn11}) and
$\lambda_{\bar{\nu}_e}$ and $\lambda_{e^+}$ are the rates of the
forward and backward reactions in Eq. (\ref{nnn12}). Since the
value of  $\sum_i N_i A_i$ does not change with weak (neutrino) 
interactions, one can
rewrite Eq. (\ref{nnn20}) in terms of mass fractions
\begin{equation}
\label{nnn21}
\frac{dX_p}{dt} = - ( \lambda_{\bar{\nu}_e} + \lambda_{e^-} ) X_p
+  ( \lambda_{\nu_e} + \lambda_{e^+} ) X_n .
\end{equation}
In the absence of heavier nuclei one has 
\begin{equation}
\label{nnn22}
Y_e = X_p + \frac{1}{2} X_{\alpha} .
\end{equation}
Because of the very large binding
energy of alpha particles the rate of its interactions with
neutrinos is nearly zero and we can write $dY_e /dt = dX_p /dt$.
Using the constraint $X_p + X_n + X_{\alpha} = 1$ and Eq.
(\ref{nnn22}), Eq. (\ref{nnn21}) can be rewritten as
\begin{equation}
\label{nnn23}
\frac{dY_e}{dt} = \lambda_n - ( \lambda_p + \lambda_n) Y_e
+ \frac{1}{2} ( \lambda_p - \lambda_n ) X_{\alpha},
\end{equation}
where we introduced the total proton loss rate $\lambda_p =
\lambda_{\bar{\nu}_e} + \lambda_{e^-}$ and the total neutron loss
rate $\lambda_n = \lambda_{\nu_e} + \lambda_{e^+}$.  If the environment reaches an equilibrium with respect to the weak interactions, 
$Y_e$ stops  changing: $dY_e /dt =
0$. From Eq. (\ref{nnn23}) one can write the equilibrium value of the
electron fraction
\begin{equation}
\label{nnn24}
Y_e = \frac{\lambda_n}{\lambda_p + \lambda_n}
+ \frac{1}{2} \frac{\lambda_p - \lambda_n}{\lambda_p + \lambda_n}
X_{\alpha} .
\end{equation}
The reaction rates in Eq. (\ref{nnn24}) are functions of the electron neutrino and electron antineutrino fluxes that reach the site of nucleosynthesis. These fluxes are in turn controlled by either collective or matter-enhanced oscillations of neutrinos between their origin and the nucleosynthesis site. Neutrino properties need to be taken into account in understanding the formation and distribution of the elements in the Universe. For further details we refer the reader to the literature (see, e.g. Ref. \cite{Volpe:2014yqa}). 

\section{Conclusions}

Symmetry properties and algebraic approaches can play a significant role in describing neutrino propagation with or without a background of other particles, including neutrinos. Appropriate techniques and relevant results are usually scattered throughout the literature: this article 
brings together several such techniques and results. In particular, the utility of symmetries is illustrated with examples chosen from the see-saw mechanism and both matter-enhanced and collective neutrino oscillations.

\appendix
\section{Gaudin Method}

In this appendix we summarize the method Gaudin introduced to study spin Hamiltonians \cite{gaudin}. One starts with the following algebra: 
\begin{equation}
\label{c1}
[S^+(\lambda),S^-(\mu)]=2\frac{S^0(\lambda)-S^0(\mu)}{\lambda-\mu},
\end{equation}
\begin{equation}
\label{c2}
[S^0(\lambda),S^{\pm}(\mu)]=\pm\frac{S^{\pm}(\lambda)-S^{\pm}(\mu)}
                                                  {\lambda-\mu},
\end{equation}
\begin{equation}
\label{c3}
[S^0(\lambda),S^0(\mu)]=[S^{\pm}(\lambda),S^{\pm}(\mu)]=0 .
\end{equation}
In the above equations $\lambda$ and $\mu$ are arbitrary complex parameters.  
Considering mutually commuting SU(2) algebras:
\[
[\hat{J}^+_i, \hat{J}^-_j ] = 2 \delta_{ij} \hat{J}^0_j, \>\>\>\>\>\>\>
[\hat{J}^0_i, \hat{J}^{\pm}_j] = \pm \delta_{ij} \hat{J}^{\pm}_j .
\]
a realization of the Gaudin algebra can be given as: 
\begin{equation}
\label{gaudinwithsu2}
S^{0}(\lambda)=A+\sum_{k}\frac{\hat{J}^0_k}{\omega_k-\lambda} 
\quad\mbox{and}\quad
S^{\pm}(\lambda)=\sum_{k} \frac{\hat{J}^{\pm}_k}{\omega_k-\lambda} ,
\end{equation}
where $\omega_k$ and $A$ are arbitrary constants. For applications to the collective neutrino oscillations we choose $\omega_p=\delta m^2/2p$. Note that
\begin{equation}
[\hat{J}^0,S^{\pm}(\lambda) ] = \pm S^{\pm}(\lambda) 
\end{equation}
where $\hat{J}^0 = \sum_i \hat{J}^0_i$. 
The operators 
\begin{equation}
\label{c4}
X(\lambda)=S^0(\lambda)S^0(\lambda)+\frac{1}{2}S^+(\lambda)S^-(\lambda)+
           \frac{1}{2}S^-(\lambda)S^+(\lambda)
\end{equation}
commute for different values of the parameters:
\begin{equation}
[X(\lambda),X(\mu)]=0  , \>\>\> \lambda \neq \mu. 
\end{equation} 
One also gets 
\begin{equation}
[X(\lambda), \hat{J}^0 ] = 0. 
\end{equation}
A lowest weight vector $| 0 \rangle$ is chosen to satisfy the conditions 
\begin{equation}
S^-(\lambda)|0 \rangle =0,\quad\mbox{and}\quad
S^0(\lambda)|0 \rangle =W(\lambda)|0 \rangle ,
\end{equation}
indicating that the state $|0 \rangle$ is an eigenstate of the operator $X(\lambda)$: 
\begin{equation}
X(\lambda)|0 \rangle = \left[ W(\lambda)^2-W'(\lambda) \right] |0 \rangle ,
\end{equation}
where prime denotes derivative with respect to $\lambda$. 

To find other eigenstates of the operator in Eq. (\ref{c4}) we consider the state $|\xi \rangle 
\equiv S^+(\xi)|0 \rangle$ for an arbitrary complex number 
$\xi$. One gets 
\begin{equation}
[X(\lambda),S^+(\xi)]=\frac{2}{\lambda-\xi}\left(S^+(\lambda)S^0(\xi)
-S^+(\xi)S^0(\lambda)\right).
\end{equation}
Hence, if $W(\xi)=0$, then $S^+(\xi)|0 \rangle$ is an eigenstate of
$X(\lambda)$ with the eigenvalue
\begin{equation}
E_1(\lambda)= \left[ W(\lambda)^2-W'(\lambda) \right] 
-2\frac{W(\lambda)}{\lambda-\xi}.
\end{equation}
This procedure can be generalized. Indeed a state of the form 
\begin{equation}
\label{unnorstate}
|\xi> \equiv
|\xi_1,\xi_2,\dots,\xi_n>\equiv S^+(\xi_1) S^+(\xi_2)\dots
S^+(\xi_n)|0>
\end{equation}
is an eigenvector of $H(\lambda)$ if the complex numbers
$\xi_1,\xi_2,\dots, \xi_n$ satisfy the so-called Bethe 
Ansatz equations:
\begin{equation}
\label{Ba}
W(\xi_\alpha)=\sum_{ {\beta=1}\atop{(\beta\neq\alpha)} }^n
\frac{1}{\xi_\alpha-\xi_\beta} \quad \mbox{for} \quad
\alpha=1,2,\dots,n.
\end{equation}
Corresponding eigenvalue is
\begin{equation}
\label{gaudineigenvalue}
E_n(\lambda) = \left[ W(\lambda)^2-W'(\lambda) \right] 
-2\sum_{\alpha=1}^n
\frac{W(\lambda)-W(\xi_\alpha)}{\lambda-\xi_\alpha}.
\end{equation}
The state in Eq. (\ref{unnorstate}) is not normalized. The normalized eigenstate can be formally written as 
\begin{equation}
\label{norstate}
|\xi> = 
|\xi_1,\xi_2,\dots,\xi_n> = Q^+(\xi_1) Q^+(\xi_2)\dots
Q^+(\xi_n)|0>
\end{equation}
where we defined 
\begin{equation}
Q^+ (\lambda) = S^+ (\lambda) \frac{1}{\sqrt{S^-(\lambda) S^+(\lambda)}} .
\end{equation}
This definition works for all states except for the highest weight state. 

At this point we adopt the notation $\mathbf{J} (\mathbf{p}) = \mathbf{J}_p$. 
To establish the connection to the neutrino Hamiltonians we write $X(\lambda)$ explicitly using the realization given in Eq. 
(\ref{gaudinwithsu2}):
\begin{equation}
\label{b6}
X(\lambda) = \sum_p \frac{\mathbf{J}_p^2}{(\omega_p - \lambda)^2} + \sum_{p,q,p\neq q}  \frac{\mathbf{J}_p \cdot \mathbf{J}_q}{(\omega_p-\lambda)(\omega_q-\lambda)} + 2 A \sum_p \frac{J^0_p}{(\omega_p-\lambda)} + A^2.
\end{equation}
Clearly $[\mathbf{J}^2_p, S_i({\mu})]=0=[\mathbf{J}^2_p, X(\mu)]$. Then
\begin{equation}
\label{b7}
[{\cal H}(\lambda),X(\mu)] =0
\end{equation}
and 
\begin{equation}
\label{bb7}
[{\cal H}(\lambda),{\cal H}(\mu)] = 0 
\end{equation}
where we defined
\begin{equation}
\label{b8}
{\cal H}(\lambda) = \sum_{p,q,p\neq q}  \frac{\mathbf{J}_p \cdot \mathbf{J}_q}{( \omega_p-\lambda)(\omega_q-\lambda)} + 2 A \sum_p \frac{{J}_p^0}{(\omega_p-\lambda)} . 
\end{equation}
Since 
\begin{equation}
 \frac{1}{( \omega_p-\lambda)(\omega_q-\lambda)}  = \frac{1}{(\omega_q - \omega_p)} \left( \frac{1}{\omega_p - \lambda} - \frac{1}{\omega_q - \lambda} \right), 
\end{equation}
we can rewrite Eq. (\ref{b8}) as 
\begin{equation}
\label{bbb1}
{\cal H}(\lambda) = - 2 \sum_{p,q,p\neq q}  \frac{\mathbf{J}_p \cdot \mathbf{J}_q}{( \omega_p-\lambda)(\omega_p-\omega_q)} + 2 A \sum_p \frac{{J}_p^0}{(\omega_p-\lambda)} . 
\end{equation}

Note that
\begin{equation}
\label{b9}
\lim_{\lambda \rightarrow \omega_p} (\lambda - \omega_p) {\cal H}(\lambda) = 2 \sum_{q,q\neq p} \frac{\mathbf{J}_p \cdot \mathbf{J}_q}{\omega_p - \omega_q} - 2 A {J}_p^0. 
\end{equation}
Now defining
\begin{equation}
\label{b11}
A= \frac{1}{\mu},
\end{equation}
Eq. (\ref{b9}) gives our invariants in the mass basis:
\begin{equation}
\label{b12}
\frac{2h_p}{\mu} = 2 \sum_{q,q\neq p} \frac{\mathbf{J}_p \cdot \mathbf{J}_q}{\omega_p - \omega_q} - \frac{2}{\mu}  J_p^0. 
\end{equation}
Multiplying Eq. (\ref{b12}) with $\omega_P$ and summing over $p$ in Eq. (\ref{b12}) gives the Hamiltonian of Eq. (\ref{singleangle}):
\begin{equation}
\label{b13}
\frac{H}{\mu} = \sum_p \omega_p \frac{h_p}{\mu} =  \sum_{q,p, q\neq p} \mathbf{J}_p \cdot \mathbf{J}_q 
- \frac{1}{\mu} \sum_p \omega_p J_p^0. 
\end{equation}
We now showed
\begin{equation}
\label{b15}
[h_p, \mathtt{H}] =0, 
\end{equation}
\begin{equation}
\label{b16}
[X(\lambda), \mathtt{H}] =0, 
\end{equation}\begin{equation}
\label{b17}
[X(\lambda),h_p] =0, 
\end{equation}
Clearly we can generate more invariants by taking different values of $\lambda$. 

One can calculate the eigenvalues of $h_p$ by taking the appropriate limit of Eq. (\ref{gaudineigenvalue}). Designating the eigenvalue 
of the Casimir operator ${\mathbf J}_p^2$ as $j_p(j_p+1)$, after some algebra one obtains 
\begin{equation}
\label{b24}
h_p |\xi> = \epsilon_p |\xi>
\end{equation}
where
\begin{equation}
\label{b25}
\epsilon_p =  \mu \sum_{q,q\neq p} \frac{j_pj_q}{\omega_p - \omega_q} + \frac{1}{2}  j_p - \mu j_p  \sum_{\alpha=1}^n
\frac{1}{\omega_p-\xi_\alpha}
\end{equation}
provided that $\omega_p \neq \xi_\alpha$. For the eigenvalues of the Hamiltonian in  Eq. (\ref{singleangle}) we then get 
\begin{equation}
E_n = \mu \sum_{p \neq q} j_pj_q + \frac{1}{2} \sum_p \omega_p j_p - \mu n \sum_p j_p + \mu \frac{n(n-1)}{2} - \frac{1}{2} \sum_{\alpha} \xi_{\alpha}. 
\end{equation}
Since the Hamiltonian is Hermitian, these eigenvalues must be real, indicating that $\xi_{\alpha}$ are either all real or come in complex conjugate pairs.

\ack

This work was supported in part
by the U.S. National Science Foundation Grants No. PHY-1514695 and PHY-1806368.

\vskip 2cm


\end{document}